\documentclass[twocolumn,preprintnumbers,amsmath,amssymb,nofootinbib
,superscriptaddress]{revtex4}
\usepackage{hyperref}
\usepackage{graphicx}
\usepackage{color}

\newcommand{\be}{\begin{equation}}  
\newcommand{\ee}{\end{equation}}  
\newcommand{\bear}{\begin{eqnarray}}  
\newcommand{\eear}{\end{eqnarray}}  
\newcommand{\ba}{\begin{array}}  
\newcommand{\ea}{\end{array}}

\newskip\humongous \humongous=0pt plus 1000pt minus 1000pt

\newif\ifdtup

\def\oldreffmt#1{\rlap{[#1]} \hbox to 2\parindent{}}

\def\figfmt#1{\rlap{Figure {#1}} \hbox to 1in{}}  
  
\def\ie{\hbox{\it i.e.}{}}	  
\def\eg{\hbox{\it e.g.}{}}	  
\def\etal{\hbox{\it et al.}}  
  
\def\beq{\begin{equation}}  
\def\eeq{\end{equation}}  
\def\bea{\begin{eqnarray}}  
\def\eea{\end{eqnarray}}

\def\bq{\begin{quote}}  
\def\eq{\end{quote}}

\def \etal {{\it et al.}\ }  
\relax  

\newdimen\tdim  
\tdim=\unitlength  
\def\bar{\overline}

\begin{document}

\preprint{FERMILAB-PUB-15-105-T}

{\title{Axion Induced Oscillating Electric Dipole Moments}

\author{Christopher T. Hill}
\email{hill@fnal.gov}
\affiliation{Fermi National Accelerator Laboratory\\
P.O. Box 500, Batavia, Illinois 60510, USA\\$ $}

\date{\today}

\begin{abstract}
The axion electromagnetic anomaly induces an oscillating electric dipole 
for {\em any} static magnetic dipole. 
Static electric dipoles do not produce
oscillating magnetic moments.  This is a low energy theorem which is a consequence of
 the space-time dependent cosmic background
field of the axion in the limit that it is only locally time dependent $(\overrightarrow{\beta}=0)$.  The electron will acquire
an oscillating electric dipole of frequency $m_a$ and strength $\sim 10^{-32}$ e-cm,
 two orders of magnitude above the nucleon, and 
within four orders of magnitude of the present standard model DC limit.   This may
suggest sensitive new experimental venues for the axion dark matter search.

\end{abstract}


\maketitle

\section{Introduction}

The axion is a hypothetical, low--mass pseudo-Nambu-Goldstone boson (PNGB) that offers 
a  solution to the strong CP problem of the standard model, and
simultaneously  provides a compelling dark matter candidate.
The expected mass scale of the axion is $m_{a} \approx {m_{\pi }^{2}}/{f_A}$ where
typical expected values of the decay constant $f_A$ range from $\sim 10^{10} $ GeV upwards
\cite{PQ1,Weinberg1,Wilczek1}.    

The axion is  expected to have an anomalous coupling to the electromagnetic field $
\overrightarrow{E}\cdot $ $\overrightarrow{B},$ taking the form:
\beq
\frac{g_{A\gamma\gamma}}{4}\left( \frac{a}{f_A}\right) F_{\mu\nu}\widetilde{F}^{\mu\nu}=
-g_{A\gamma\gamma}\left( \frac{a}{f_A}\right) \overrightarrow{E}
\cdot \overrightarrow{B}
\eeq
where $\widetilde{F}_{\mu\nu}=(1/2)\epsilon_{\mu\nu\rho\sigma}{F}^{\rho\sigma}$, 
and  $g_{A\gamma\gamma} $  is the dimensionless
anomaly coefficient. In various models we have \cite{PDG}: 
\bea
g_{A\gamma\gamma} & \approx & 8.3\times 10^{-4}\qquad \;\;\; \makebox{DFSZ \cite{DFSZ}}
\nonumber \\
g_{A\gamma\gamma} &\approx & -2.3\times 10^{-3}\qquad \makebox{KSVZ \cite{KSVZ}}
\eea
We will quote results below, scaled by $g_{A\gamma\gamma}/10^{-3}$.

Most strategies for detecting the cosmic axion
exploit the electromagnetic anomaly \cite{Bj,Sikivie1,Graham}
together with the assumption of a coherent galactic dark-matter background field \cite{Ipser},  
$ {a}/{f_A}\equiv \theta(t) =$\ $\theta_{0}\cos (m_at)$.  Saturating the local galactic dark-matter density  of $\sim 0.3$ GeV/cm$^{3}$ \ yields \ $\theta_0 \approx 3.7\times 10^{-19}$ \cite{Graham}.
In typical RF cavity experiments such as ADMX, one applies a large external constant magnetic field to the cavity,
$\overrightarrow{B}_0$ and the
anomalous coupling to $\theta(t)$ induces an oscillating electromagnetic response field,
$\overrightarrow{E}_r$ and $\overrightarrow{B}_r$.
Combined with the conducting boundary conditions of the detector, the ``cavity modes'' can become excited,
which can  generate a resonant signal in the cavity. This
offers the possibility of both detecting the existence of the axion
and simultaneously establishing that it is a significant component of dark-matter.

Recently several authors have considered the possibility of observing an {\em oscillating
electric dipole moment for the nucleons}  \cite{budkher, graham2}. Indeed, it was the original problematic strong CP-violating nucleon
electric dipole moment, arising from  the $\theta$-term in QCD, that the axion was designed 
to cure. Nonetheless, the relic
small cosmological oscillations of the axion in its potential about zero,  $\theta(t)$, will
induce a small oscillating electric dipole moment for the nucleons with a frequency $m_a$
given by $d_N\sim 10^{-16}\theta(t)\approx 3.67\times 10^{-35}\cos(m_at)$ e-cm \cite{budkher}.  
Thus far, this effect has only been considered to be specific to
baryons. It arises, not by the electromagnetic anomaly $\propto g_{A\gamma\gamma}$,  but rather directly
via the QCD-induced axion potential.

In the present paper we show that the axion electromagnetic anomaly, 
together with  cosmic axions in the $\overrightarrow{\beta}=0$ limit, leads to
a ``low energy theorem:''  {\em any
magnetic dipole source (eg, magnetization, spin magnetic moments, orbital magnetic moment,
etc.)  will develop a small, time-oscillating, CP-violating, effective electric dipole}.
Remarkably, the dual is not true:
A static electric electric polarization will {\em not} lead to a small
time-oscillating effective dipole magnetic field.  

In particular, the electron
will develop an effective oscillating electric dipole moment
proportional to  the magnetic moment, $d_e\approx 2g_{A\gamma\gamma}\theta_0\cos(m_at)\mu_{Bohr}
\approx 1.4\times 10^{-32}(g_{A\gamma\gamma}/10^{-3})\cos(m_at)$
e-cm. We use the term, ``effective,'' because this arises at the level of a one-particle reducible
Feynman diagram.  The result is intrinsically non-relativistic.

The magnetic moment of the electron is much larger
than that of the nucleon, and hence 
the axion-induced oscillating electric dipole moment is almost three orders
of magnitude larger than that of the nucleon.
Since the current best limit upon any DC elementary particle EDM is that  of the electron, 
 of order $d_e\leq 8.7\times 10^{-29}$ e-cm, \cite{ACME}, 
the electron may be a promising place
to search for an oscillating EDM.

\section{ Feynman Diagram Analysis of Induced Electric Dipole Moment}

\begin{figure}[t]
\vspace{4.5cm}
\includegraphics{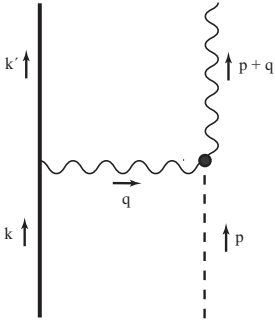}
\vspace{-0.0cm}
\caption[]{ Feynman diagram for axion induced electric dipole moment. Photon $q$ emitted
from electron magnetic moment. Solid
dot is the axion anomaly interaction, $\theta\overrightarrow{E}\cdot\overrightarrow{B}$. Dashed line: incoming
axion $\theta$, $p$. Outgoing photon: $p+q$, $\epsilon_\mu$.
Solid line: incoming electron $k$; recoil electron, $k'$. }
\label{xroots}
\end{figure}

We consider the  tree-level perturbative correction
of the axion anomaly to the electron magnetic dipole operator. 
We begin by writing the axion anomaly in terms of
vector potentials and integrating by parts:
\beq
\label{3}
\frac{1}{2}g_{A\gamma\gamma}\int d^4x \;\epsilon_{\mu\nu\rho\sigma}\partial^\mu\theta(x) 
A^\nu (x)\partial^\rho A^\sigma(x)
\eeq
where $\theta(x) = a/f_A=\theta_0\cos(m_at)$.
Likewise, we have the  Dirac operator of the magnetic moment
of the electron:
\bea
\label{mm}
&& \frac{ie}{2m_e}\int d^{4}x\; \overline{\psi }(x)\sigma _{\alpha \beta }\psi (x)\; \partial
^{\alpha }A^{\beta }(x)
\eea
where $\sigma_{\alpha \beta }=(i/2)[\gamma_\alpha,\gamma_\beta]$ 
 (we follow conventions of Bjorken and Drell \cite{BjDrell}).

The Feynman amplitude for the first order $g_{A\gamma\gamma}\theta_0$ correction to the magnetic
dipole moment follows from the time-ordered product of eqs.(\ref{3},\ref{mm})
and is the tree diagram of Fig(1).  In momentum space we have:
\bea
\label{feyn1}
&& \!\!\!\! \!\!\!\!\!\!\!\!\!\!
\frac{ie}{2m_e}\;g_{A\gamma\gamma}\theta_0\; p^\mu \epsilon^\nu (p+q)^\rho  \epsilon_{\mu\nu\rho\sigma}\times
\nonumber \\
&& \qquad \frac{1}{q^2}
\left(g^{\sigma\tau}-\lambda \frac{q^{\sigma}q^{\tau}}{q^2}  \right) q^\omega \bar{u}(k')\sigma_{\tau\omega}u(k)
\eea
where $\epsilon^\nu $ is the outgoing photon polarization, $u(k)$ is
an electron spinor, and momenta are as in Fig.(1).
Note that a factor of $2$ arises from the two possible photon field contractions.
The gauge dependent terms $\propto \lambda$ vanish due to the antisymmetry of
$\sigma_{\mu\nu}$.

The dipole moments are defined by going to the electron
rest-frame $k=(m_e, \overrightarrow{0})$. The electron is 
very heavy compared to the axion, and is therefore essentially
stationary, and absorbs $3$-momentum but not energy (zero recoil).
We assume in the electron rest frame that
the axion field momentum is approximately pure timelike, $p_\mu =(m_a, \overrightarrow{0})$
and $|\overrightarrow{q}|\approx m_a<< m_e$,  $k_0\approx k_0'$, and
the exchange photon momentum is therefore spacelike, $q=(0,\overrightarrow{q}) $.  

Since $p^\mu$ is  
timelike, we have  $p^\mu\epsilon_{\mu\nu\rho\sigma}=m_a\epsilon_{0\nu\rho\sigma}$
and we'll pass to $D=3$ latin spatial indices, 
$\epsilon_{0\nu\rho\sigma}\rightarrow \epsilon_{ijk}$.
Also,
 $\bar{u}(k')\sigma_{\tau\omega}u(k)\rightarrow \epsilon_{ij k}\chi^\dagger \sigma_k \chi$
( $(\tau\omega) \leftrightarrow (ij)$ are likewise spatial in the nonrelativistic limit).
The Dirac four-component spinor
 $\psi$ has been replaced by the two-component Pauli spinor, $\chi$, with Pauli matrices $\sigma^k $.
 The amplitude becomes:
\bea
&& \!\!\!\! \!\!\!\!\!\!\!\!\!\!
-\frac{ie}{2m_e}g_{A\gamma\gamma}\theta_0 \; m_a \; \frac{1}{\overrightarrow{q}^2}\epsilon^i q^j q^l \epsilon_{ijk}
\epsilon^{klm}\chi^\dagger \sigma_m \chi
\nonumber \\
&& =g_{A\gamma\gamma}\theta_0\; \mu_{Bohr}
\chi^\dagger \sigma_i \chi \cdot m_a \epsilon^i
\eea
where $\mu_{Bohr}=e/2m_e $. 
The photon propagator pole, $1/\overrightarrow{q}^2$ has cancelled against $\overrightarrow{q}^2$ in the numerator,
yielding a contact term.
Here we've used the identity, $\epsilon_{ijk}\epsilon^{klm}=\delta^{l}_{i}\delta^{m}_{j}-\delta^{l}_{j}\delta^{l}_{i}$.
The polarization of the outgoing photon is spacelike and transverse, $0=(p+q)\cdot\epsilon =\overrightarrow{q}
\cdot\overrightarrow{\epsilon}$. 
In Coulomb gauge with vector potential  $\overrightarrow{A}$, the electric field
is given by $\overrightarrow{E} =-\partial_t\overrightarrow{A}=  m_a \overrightarrow{\epsilon}$. Our final result can be written as an effective interaction for the non-relativistic electron as:
\bea
\label{edm}
&& \int d^4x \;2g_{A\gamma\gamma}\theta(t)\; \mu_{Bohr}\;
\chi^\dagger \frac{\overrightarrow{\sigma}}{2} \chi(x) \cdot \overrightarrow{E}(x,t)
\eea
We thus see that the  magnetic moment of the 
electron,
in the presence of the axion, yields an
oscillating electric dipole moment 
of frequency $m_a$. 
This result can be obtained in the Pauli-Schoedinger
non-relativistic spin-$1/2$ theory, or in Georgi's heavy-quark formalism applied to the ``heavy
electron'' \cite{Georgi}\cite{cth}.  

The magnitude of the electric dipole
moment is $g_{A\gamma\gamma}\theta_0 \overrightarrow{m}$ where
$\overrightarrow{m} = g\mu_{Bohr}\overrightarrow{s}$ for spin $\overrightarrow{s}$ and $g=2$.
In an applied oscillating, phase matched,  external electric field, $\overrightarrow{E}(t)\propto cos(m_a t)$,
the term eq.(\ref{edm}) in the action generates small energy 
shifts which may be in principle detectable, as in \eg, \cite{ACME}.

In general, a nonrelativistic, static
electric dipole moment   $ \overrightarrow{P}(x)\sim 2g_{A\gamma\gamma}\; \mu_{Bohr}\;
\chi^\dagger \frac{\overrightarrow{\sigma}}{2} \chi  $, modulated by $\theta(t)$,  includes a
nonlocal term: 
\beq
\small
S^{\prime }=\int d^{4}x\; \theta (t)\left( \overrightarrow{P}\cdot 
\overrightarrow{E}+\overrightarrow{\nabla }\cdot \overrightarrow{P}\left( 
\frac{1}{\overrightarrow{\nabla }^{2}}\right) \overrightarrow{\nabla }\cdot 
\overrightarrow{E}\right) 
\eeq
In an arbitrary gauge,
$\overrightarrow{E}=\overrightarrow{\nabla }\varphi
-\partial _{t}\overrightarrow{A}$ , we see that:
\bea
&& S^{\prime }  =\int d^{4}x\; \theta (t)\overrightarrow{\nabla }\cdot (
\overrightarrow{P}\varphi) 
\nonumber \\
& &\!\!\!\!\!\!\!\!  +\int d^{4}x\; \partial _{t}\theta
(t)\left( \overrightarrow{P}\cdot \overrightarrow{A}+\overrightarrow{
\nabla }\cdot \overrightarrow{P}\left( \frac{1}{\overrightarrow{\nabla }^{2}}
\right) \overrightarrow{\nabla }\cdot \overrightarrow{A}\right) 
\eea
 $S^{\prime }$ is a total divergence in the
limit $ \partial _{t}\theta (t)\rightarrow 0.$  
and is indistinguishable
from eq.\ref{edm}  when  $\overrightarrow{\nabla }
\cdot \overrightarrow{E}=0$.

Our result can be inferred from the classical Maxwell equations.  For response fields, $\overrightarrow{E }_r$ and $\overrightarrow{B}_r$, in
the presence of a constant-in-time external magnetic dipole field $\overrightarrow{B}_{0}(\overrightarrow{x})$,
we have 
$\overrightarrow{\nabla }\cdot \overrightarrow{B}_{r}=\overrightarrow{\nabla 
}\cdot \overrightarrow{E}_{r}=0$,    and:
\bea 
\label{eight}
&& \!\!\!\!\!\!\!  \overrightarrow{\nabla }\times \overrightarrow{B}%
_{r}-\partial _{t}\overrightarrow{E}_{r}=-g_{a\gamma\gamma}\overrightarrow{B}_{0}(\overrightarrow{x})
\;\partial _{t}{\theta(t) } 
\nonumber \\
&& \!\!\!\!\!\!\! \overrightarrow{\nabla }\times \overrightarrow{E}%
_{r}+\partial _{t}\overrightarrow{B}_{r}=0 
\eea
$\overrightarrow{E}_{r}(\overrightarrow{x},t)= g_{a\gamma\gamma}{\theta(t)}\overrightarrow{B}_{0}(\overrightarrow{x})$
and $\overrightarrow{B}_{r}(\overrightarrow{x},t)=0$ would be exact 
if $\overrightarrow{\nabla }\times \overrightarrow{B}_0=
\overrightarrow{\nabla }\times\overrightarrow{m}=0$. 
However,
$\overrightarrow{B}_{0}(\overrightarrow{x},t)$ contains a nonzero $\delta$-function singularity $\propto \overrightarrow{m}$.
The full radiative solution is readily obtained with retarded Green's functions, and contains $\overrightarrow{E}_{r}(\overrightarrow{x},t)= g_{a\gamma\gamma}{\theta(t)}\overrightarrow{B}_{0}(\overrightarrow{x})$ in the near-zone.  The radiated power,
representing conversion of vacuum axions to photons by the
oscillating electric dipole, is then of order
$\sim (m_a^2g_{a\gamma\gamma}{\theta}_{0}\mu_{Bohr})^2$ and is negligible \cite{cth}.

\section{Axionic Electromagnetic Duality}
 
The result of the previous section is a general low energy theorem.
Consider an arbitrary localized magnetic dipole interaction in Coulomb gauge
${\nabla}\cdot\overrightarrow{A}=0 $, $A_0=0$: 
\beq
\int d^4x \; \overrightarrow{M}(\overrightarrow{x})\cdot\overrightarrow{B} =
\int d^4x \; \overrightarrow{M}\cdot\overrightarrow{\nabla}\times\overrightarrow{A}
\eeq
The effect of the axion anomaly, to first order
in perturbation theory as in the previous section,  schematically produces a term,
\bea
& & 
-g_{a\gamma\gamma}\int d^4x\; \overrightarrow{M}(\overrightarrow{x})\cdot\overrightarrow{\nabla}\times\left( \frac{1}{\nabla^2} 
(\partial_t\theta(t))\overrightarrow{\nabla}\times\overrightarrow{A} \right)
\nonumber \\
& & = g_{a\gamma\gamma}\int d^4x \; \theta(t)\;\overrightarrow{M}\cdot\overrightarrow{E}
\eea
where ${1}/{\nabla^2} $ is shorthand for the static potential (the time averaged Feynman propagator;
presently we use the notation of ref.\cite{Deser}).
Throughout, we've assumed  that
$\theta(t)$ only has  time dependence, \ie,  $\overrightarrow{\beta}=0$. We've integrated by parts and used 
$\overrightarrow{E}=-\partial_t\overrightarrow{A}$ (in the static limit for $\overrightarrow{M}(\overrightarrow{x})$
the produced electric field $\overrightarrow{E}$ inherits  the time dependence
of $\theta(t)$).
This establishes the result in general for any magnetic dipole $\overrightarrow{M}$
acquiring an electric dipole $g_{a\gamma\gamma}\theta_0\overrightarrow{M}$. The crucial element
is the static Green's function ${1}/{\nabla^2} $ which is  enforced by the zero recoil limit.

This result is formally related to duality in electromagnetic theory.
Deser and Teitelboim \cite{Deser}  elegantly formulated the continuous 
electromagnetic dual tranformation,
whereby $\overrightarrow{E}\leftrightarrow -\overrightarrow{B}$.
This
arises from an  infinitesimal non-local transformation at the level of the vector potential.
In Coulomb gauge the Deser-Teitelboim dual transformation is:
\beq
\delta \overrightarrow{A}=\frac{\epsilon}{\nabla^2}\overrightarrow{\nabla}\times\partial_t\overrightarrow{A}
\eeq
which implies at the field strength level,
\beq
\delta \overrightarrow{E}=-\epsilon \delta \overrightarrow{B}
\qquad
\delta \overrightarrow{B}=\epsilon \delta \overrightarrow{B}
\eeq
where  $\overrightarrow{E}=-\partial_t\overrightarrow{A}$ 
and $\overrightarrow{B}=\overrightarrow{\nabla}\times\overrightarrow{A}$.
The transformation on $\delta \overrightarrow{E}$ uses the on-shell condition,
$\partial^2_t\overrightarrow{A}-\nabla^2\overrightarrow{A}=0$.   

We see that the transformation acting on the magnetic source 
term $\overrightarrow{M}\cdot\overrightarrow{\nabla}\times\overrightarrow{A}$
will produce a dual rotation of the magnetic field
into the electric field, provided we can replace the dual rotation
angle by $\epsilon \rightarrow g_{a\gamma\gamma}\theta(t)$.
\beq
\delta \overrightarrow{A}=\frac{g_{a\gamma\gamma}\theta(t)}{\nabla^2}\overrightarrow{\nabla}\times\partial_t\overrightarrow{A}
\eeq
Can we promote the static Deser-Teitelboim transformation to
a time dependent transformation?

We might worry that this
affects the kinetic term of the electromagnetic theory
(note that the $g_{a\gamma\gamma}\theta(t)\overrightarrow{E}\cdot\overrightarrow{B}$ term is already infinitesimal
in this sense and does not transform).
However,  we can see that the electromagnetic action,
$S=\int d^4x(\overrightarrow{E}^2-\overrightarrow{B}^2)/2$,  is invariant under
the time dependent transformation.  Define  $\epsilon(t) = g_{a\gamma\gamma}\theta(t) $
and consider:
\bea
\label{EE}
\delta \overrightarrow{A} & = &\frac{1}{\nabla ^{2}}\epsilon (t)\left( \nabla
\times \partial _{t}\overrightarrow{A}\right) 
\nonumber \\
\delta \overrightarrow{E} & = &
-\left( \partial _{t}\epsilon
\right) \frac{1}{\nabla ^{2}}\left( \nabla \times \partial _{t}%
\overrightarrow{A}\right) -\epsilon \overrightarrow{B}
\nonumber \\
\delta \overrightarrow{B}& =& \epsilon \frac{1}{\nabla ^{2}}\overrightarrow{%
\nabla }\times \left( \nabla \times \partial _{t}\overrightarrow{A}\right)
=\epsilon (t)\overrightarrow{E}\qquad 
\eea
where we follow \cite{Deser} and use the vector potential equation of motion, 
$\partial_t^2\overrightarrow{A} = \nabla ^{2}\overrightarrow{A}$.
We see that $\delta \overrightarrow{B}$
has the same form as that of Deser and Teitelboim. Hence the magnetic dipole moment will
cleanly rotate into an oscillating electric dipole moment.

If we consider the action integral we find:
\bea
\int \frac{1}{2}\delta \overrightarrow{E}^{2}& = -&\int \epsilon \overrightarrow{E}\cdot \overrightarrow{B}+\int \partial
_{t}\epsilon \overrightarrow{A}\cdot \left( \nabla \times \overrightarrow{%
A}\right)
\nonumber \\
\int \frac{1}{2}\delta \overrightarrow{B}^{2} & = &\int \epsilon 
\overrightarrow{E}\cdot \overrightarrow{B}
\eea
where we've integrated by parts in space and time and discarded surface terms.
Note that, with some manipulation, $\int \epsilon \overrightarrow{E}\cdot 
\overrightarrow{B}=\frac{1}{2}\int \partial _{t}\epsilon $ $%
\overrightarrow{A}\cdot \nabla \times \overrightarrow{A}$ + total divergence.
We thus find for the shift in the action:
\bea
\label{16}
\delta S & =& -2\int \epsilon \left( \overrightarrow{E}\cdot 
\overrightarrow{B}\right) +\int \partial _{t}\epsilon  \overrightarrow{A}%
\cdot \left( \nabla \times \overrightarrow{A}\right) 
\nonumber \\
& = & 0 
\eea
modulo surface terms.

The remarkable identical cancellation in eq.(\ref{16}) occurs because of the 
 time dependent $\epsilon(t)$.  Essentially the $\epsilon(t) \overrightarrow{E}\cdot \overrightarrow{B} $ 
term becomes physical when $\epsilon(t)$ has time dependence (\ie, if $\epsilon$ is a constant
then the Deser-Teitelboim transformation applied to the action is
$\delta S  = -2\int \epsilon  \overrightarrow{E}\cdot 
\overrightarrow{B} $  which is a total divergence). The physical axion-$\gamma\gamma$ interaction cannot be
rotated away by the dual transformation.   Hence the physical dual rotation induced
by the axion in Fig.(1)
affects only the source terms. We can view the induced electric dipole moment as 
a physical, duality induced rotation of the magnetic source, via the axion.

However, static electric dipole moments {\em will not} acquire  oscillating magnetic moments.
The  effect of
the additional nonlocal term in eq.(\ref{EE}) of  $\delta \overrightarrow{E}$  is nontrivial.  Given an electric
dipole moment term
in the action,  $\int d^4x\;\overrightarrow{P}\cdot\overrightarrow{E}$ where $\overrightarrow{P}$ is time
independent, 
we find $\delta \int d^4x\;\overrightarrow{P}\cdot\overrightarrow{E}=0$ upon integrating the
nonlocal term in $\delta \overrightarrow{E}$ by parts and using 
$\partial^2_t\overrightarrow{A}-\nabla^2\overrightarrow{A}=0$. 
The asymmetry between magnetic and electric dipoles in axion electrodynamics is a
 consequence of the exclusive time dependence in $\epsilon(t)$ (this is modified if 
$\beta_{axion} \neq 0$).

If we introduce large classical magnetic background fields, the physical dual rotation induced
by the axion on these fields generates the solutions 
to Maxwell's equations. 
In an RF cavity experiment with a
large applied constant magnetic field $\overrightarrow{B}_{0}=B_0\hat{z}$  the Maxwell equations
take the form of eq.(\ref{eight}) and above. 
The {rhs} of eq.(\ref{eight}) is just the time derivative
of the dual rotation of the large applied field 
$B_0$. The particular solution of 
is likewise the infinitesimal dual tranformation of $\overrightarrow{B}_0$, 
 $ \overrightarrow{E}_r = -g_{a\gamma\gamma}\theta_0(t)\overrightarrow{B}_{0}$.

In  a cylindrical cavity we will also have the homogeneous solution, \eg,  the
lowest transverse magnetic mode with frequency $m_a$:  $ \overrightarrow{E}_h =  k\theta_0(t)J_0(m_a r)\hat{z}$
and $ \overrightarrow{B}_h =  k\theta_0(t)J_1(m_a r)\hat{z}$.  The conducting boundary condition at the cavity
wall, $r=R$, implies $ \overrightarrow{E}(R) =\overrightarrow{E}_r+\overrightarrow{E}_h(R)= 0$.
 This condition locks the arbitrary
constant, $k$, to  $g_{a\gamma\gamma}$. This, in turn, leads to resonance if the
cavity is designed so that $J_0(m_a R)\rightarrow 0$.

\section{Conclusions}

We have obtained an induced 
oscillating  electric dipole moment for the electron,
proportional to  the magnetic moment, $2g_{A\gamma\gamma}\theta_0\cos(m_at)\mu_{Bohr}$.
The result is quantitatively $
\approx 1.4\times 10^{-32}(g_{A\gamma\gamma}/10^{-3})\cos(m_at)$
e-cm. 
The result is  two orders of magnitude greater
than the typical result expected
for the nucleon,  
$d_N\sim  3.67\times 10^{-35}\cos(m_at)$ e-cm \cite{budkher}, 
and within four orders of magnitude
of the DC limit on the EDM  of the electron, 
 $d_e\leq 8.7\times 10^{-29}$ e-cm, \cite{ACME}. 

The result is a general low energy theorem and applies to any
static magnetic system.
Axion electromagnetic anomaly effects are essentially local oscillating
 dual rotations that lead to potentially observable signals.
The axion anomaly
perturbs the system by locally producing a physical, infinitesimal,
time dependent dual rotation of 
$\delta\overrightarrow{E} = g_{a\gamma\gamma}\theta_0(t)\overrightarrow{B}$,
effectively rotating a magnetic moment to an oscillating electric dipole moment.
The duality
of axion electrodynamics also implies 
the absence of induced oscillating magnetic moments from static electric dipoles.
Further technical details will be presented elsewhere \cite{cth}.

The existence of such phenomena may
imply a number of potentially sensitive venues. Existing sensitive
DC experiments utilizing molecular beams, such as the ACME experiment,
might be adapted to search for AC dipole moments \cite{ACME}.  
The anomalous induced electric dipole moment applies as well
to the nucleon, but there we expect it is only comparable to the
direct effect from nonperturbative QCD \cite{budkher}. 



\vskip 5 pt
\noindent
{\bf Acknowledgements}
 
I thank Aaron Chou, Estia Eichten, Andrew Sonnenschein, 
for useful discussions.
This work was done at Fermilab, operated by Fermi Research Alliance, 
LLC under Contract No. DE-AC02-07CH11359 with the United States Department of Energy.

\end{document}